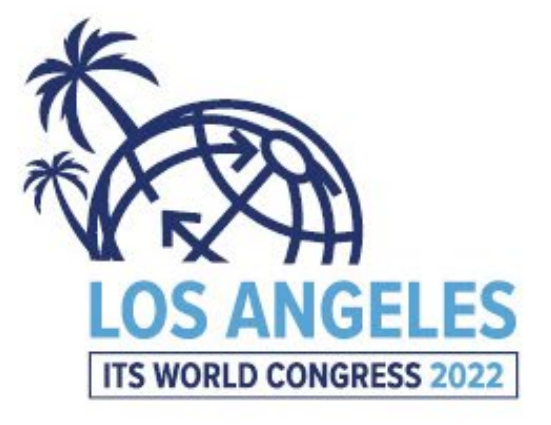

# Evaluation of a beam switching smart antenna array for use in traffic telematics V2X applications

**Hagen Ußler[1*], Oliver Michler[1]**
1. Technische Universität Dresden, Institute of Traffic Telematics, 01062 Dresden, Germany
E-Mail: {hagen.ussler, oliver.michler}@tu-dresden.de

**Abstract**
With the digitalization of transportation new use cases for digital information services are emerging. For example, prioritization of road users at traffic signals, especially emergency vehicles, is a desired goal. Requirements for the necessary communication link between road user and C-ITS station are often inadequately met. In particular, increasing communication distances while minimizing latencies are key requirements for timely influencing the control of traffic light signal programs. This paper presents the automotive application of a fast-switchable antenna array with targeted transmission direction selection and high antenna gain. In addition to antenna design requirements, theoretical analyses to increase transmission distance are performed using radio propagation simulation. Furthermore, practical evaluation is performed both in laboratory and test field environment using continuous wave and C-ITS service measurements. An automatic switching of antenna sectors based on geolocation is implemented and discussed from a scientific point of view. As a result, using an adaptive antenna array in traffic telematics environments is proposed to provide more robust communication links and increase the radio transmission distance.

**Keywords:**
V2X, Smart Antenna Beam Switching, C-ITS, Radio Propagation Simulation, Emergency vehicle

## Introduction

The development of digital and connected traffic systems, including cooperative intelligent transportation systems (C-ITS) and their traffic telematics applications, has been of great interest to various stakeholders in recent years, both on the part of service providers in the field of automated driving functions and infrastructure operators. The widespread deployment of connected vehicles and infrastructures (V2X) will make a positive contribution to increasing road safety through predictive information acquisition and transmission, reducing environmental damage through, for example, more efficient driving maneuvers, but also the general provision of comfort functions through, for example, infotainment systems. Increasing the efficiency of existing and future transport systems is therefore the focus of scientific and technical research and development for all stakeholders across all transport modes. In particular, automated and autonomous driving functions as well as the introduction of more efficient, low-latency communication infrastructures between onboard units (OBU) and roadside units (RSU) should be mentioned.
In addition to the IEEE 802.11p radio standard for vehicle ad-hoc networks, the communication standards of the fourth (LTE-V) and fifth (5G) mobile communications generation are particularly widespread for V2X applications in the transportation sector or will be used for in the future. Advantages and disadvantages of the respective standards are subject of controversial discussion in the scientific community [1]. However, which technology is used in each case often depends on the respective vehicle manufacturer and infrastructure operator. The information technology optimization approaches on which this paper is based are operator-independent and therefore fundamental research.
Against the background of service availability, the provision of a radio link between the C-ITS stations is essential in the V2X scenario. In the area of road traffic, the scenario of giving priority to emergency vehicles at traffic lights is such an application, which is also described in [2]. The vehicles advanced notification is ensured by radio communication and corresponding standardized messages, e.g. Decentralized Environmental Notification Messages (DENM). However, the time of switching signal programs for green clearance of the emergency vehicles corresponding lane is usually longer than the time until the vehicle approaches the intersection. As a result, it may not be possible to ensure free passage through the intersection. Therefore, in order to realize a timely notification at the RSU and a fast green clearance at the traffic light system during a rescue operation, the radio communication distance has to be increased. This can be realized by special antenna devices with high antenna gain, which are considered in this paper. They can be attached to the RSU on the infrastructure side or to an OBU on the vehicle side.
Usually, the equipment of the trackside infrastructures by RSUs is carried out after a radio-optimized site planning by means of radio propagation simulation in order to ensure an optimal radio propagation. This is particularly influenced by the examination environment [3]. However, in an urban environment, RSU locations are usually determined by the locations of traffic lights at intersections. This creates special challenges for robust radio communication between the RSU and the OBU of an emergency vehicle, especially when the road trajectory to the traffic light is along a curve. Therefore, the use of several fast switching directional antennas on the vehicle roof with specific beam angle is proposed, which is implemented in a special beam switching antenna apparatus.
By using a radio simulation tool, the signal propagation of such an antenna design can be simulated in a known radio environment. This allows conclusions to be drawn about the received signal level or the quality of service even before measurement. The developed beam switching antenna was adapted for the frequency range of 5.9 GHz, making it applicable for V2X communication services based on ETSI ITS-G5 [4].
This scientific paper presents an optimized antenna concept with beam-switching method using geolocation for a V2X scenario, contributing to:

- Increasing the communication distance through high antenna gain,
- Increasing radio coverage by switchable antenna sectors with high directivity,
- Ensuring targeted and fast communication through low switching latency of the antenna sectors.

The paper is structured as follows: After the introduction in Chapter 1, Chapter 2 describes the theoretical basis of radio propagation modeling and antenna characterization as well as the concept of a traffic telematics beam switching scenario. Then, Chapter 3 discusses investigations of the designed antenna array both in a laboratory environment, in a radio simulation, and in a testbed using continuous wave (CW) measurements. Chapter 4 considers an applied measurement setup for V2X services with automated beam-switching algorithm in an urban study environment. The paper concludes with Chapter 5, Summary.

## 2. Fundamentals

### *2.1 Radio propagation modeling and antenna characteristics*

For the communication links in V2X applications considered in this paper, the receive power, $P_R$ (dBm), calculated in the receive module is primarily decisive for assessing the receive quality. According to [5], this is calculated using the link budget equation to be

$$P_R = P_T - C_T + G_T - PL_{FS} - PL_{Div} + G_R - C_R \, . \qquad (1)$$

$P_T$ (dBm) represents the transmit power of the transmitter module, $C_T$ / $C_R$ (dB) the cable and connector losses, $G_T$ / $G_R$ (dBi) the antenna gain on the transmitter and receiver side, $PL_{FS}$ (dB) the path loss of the signal due to free space propagation and $PL_{Div}$ (dB) the path loss of the signal due to various propagation phenomena such as multipath effects, reflections, refractions, diffractions or scattering.

The free space loss is influenced by the signal frequency $f$ (Hz) and the distance $d$ (m) between transmitter and receiver and is calculated with the speed of light $c$ (m/s) in the signal propagation medium

$$PL_{FS} = 20 \, lg \, (d) \, + 20 \, lg \, (f) - 20 \lg(c/4\pi) \, . \qquad (2)$$

According to Eq. (1), an increase in signal receive power in a given radio environment can be achieved by a higher antenna gain $G$.

According to the inverse-square law, the received power of a signal at an isotropic receiving antenna with defined aperture, which propagates spherically from a point-like signal source with defined $P_T$, is inversely proportional to the squared radius of the spherical surface, which corresponds to the distance $d$ [6]

$$P_R = P_T \left[ \frac{c}{4\pi d f} \right]^2 . \qquad (3)$$

Accordingly, the received power is reduced by a quarter when the distance is doubled, which corresponds to a power level reduction of 6 dB. All other parameters being equal, according to Eq. (1), the antenna gain is therefore decisive for the increase in transmission distance aimed at in this work. In other words, a doubling of the transmission distance can be achieved with a 6 dB increase in the signal receive level $P_R$.

The antenna gain is calculated according to [7] from the product of directivity $D$, and the antennas radiation efficiency factor $\eta \leq 1$, characterizing electrical losses of the antenna, which may occur, e.g., due to ohmic line resistances

$$G = 10 \lg(D \cdot \eta). \qquad (4)$$

The directivity factor of a non-isotropic antenna indicates the ratio of its radiation intensity, $U$ (W/unit sold angle), in a defined direction to the radiation intensity of an isotropic radiator $i$,

$$D = \frac{U \, (\theta, \varphi)}{U_i} . \qquad (5)$$

The partial directivity is determined from the subcomponents in horizontal azimuth angle, $\theta$, and vertical elevation angle, $\varphi$, in a spherical coordinate system [7].

### *2.2 Radio propagation simulation*

The use of radio systems in a specific environment requires consideration of the local conditions. This applies in particular to the investigation of antenna systems, which often have a considerable influence on radio propagation or the received signal level due to a characteristic antenna pattern. Therefore, for the investigation of signal propagation in a known environment, radio propagation simulations are suitable. These are preferably used for planning antenna sites or for optimized infrastructure dimensioning, since a reproducible investigation of signal propagation effects and radio coverage is possible. For the use of smart antennas in the traffic environment considered in this work, the radio propagation simulation is beneficial, since the possibility of switching the antenna elements in different directions also results in differences with regard to signal propagation and signal receive levels.

In general, radio propagation algorithms can be classified into empirical, semi-empirical, numerical and ray-based methods. In empirical methods, measurements are used to determine the model parameters, while ray-based methods consider the propagation of individual signal paths for signal transmissions in particular [8]. The basis of a radio propagation simulation is the modeling of a radio environment by geometric objects and assignment of frequency-dependent electrical material properties, which have an influence on signal attenuation by, for example, reflections at their surfaces or by transmissions. Furthermore, the result of the radio propagation simulation is significantly influenced by the radio model, which specifies the relevant radio properties on the transmitting and receiving side and defines the simulation parameters. Relevant work for radio planning simulations related to traffic telematics applications can be found for example in [3] for C-ITS communication in road traffic, in [9] for V2X radio coverage using leaky coaxial cables along road crash barriers or in [10] in parking garages.

*2.3 V2X Beam Switching concept and antenna design*

Communication applications with high demands on transmission distance and coverage require a special antenna concept and design. Various smart antenna concepts are known in scientific research, classified as adaptive antennas, beam-switching antennas and beam-forming antennas [11]. Smart antennas are characterized in particular by an increase in capacity, radio coverage and distance, and a general increase in quality of service [12]. Beam-switching antennas not only have a high directivity and the associated high antenna gain when implemented as a horn antenna array, but also the advantage of enabling comparatively short switching times of the antenna elements, since there is no need for a phase shifter. This is particularly useful when implementing V2X applications with high latency requirements and in dynamic traffic environments. Switching of the antenna elements can be performed by electromechanical as well as solid-state switches. The latter have the advantage of enabling particularly fast switching processes in the nanosecond range as well as being more compact and requiring less maintenance. However, the disadvantages are lower isolation and higher power losses compared with electromechanical switches [13]. To achieve a high transmission distance, antenna elements with a high antenna gain are beneficial, which are usually implemented as horn elements or patch antennas. The small aperture angle of a horn antenna element due to the required high directivity with high antenna gain results in a small directional antenna pattern in azimuth and elevation plane. Therefore, in order to enable a larger radio coverage in the horizontal plane, it is necessary to arrange several horn antenna elements with different directional beam angles as an array. The implementation of individually switchable antenna elements has the advantage of providing the required directivity over a wider azimuth angle. A precise directional selection with a high antenna gain can thus be realized.

For the traffic scenarios considered in this paper, this is particularly advantageous for the communication link between OBU and RSU when an emergency vehicle approaches at the intersection with traffic lights. Other communication scenarios include targeted radio links while taking bends or when there are obstacles within the Line of Sight (LOS) between a transmitter and receiver.

The basic concept of communication with smart switchable antennas in the V2X scenario is shown in Fig. 1. Here, a) shows that a communication link between emergency vehicle and RSU is only possible via a directional but narrow radio beam. Signal reception using only omnidirectional antennas is not possible at this distance or in this environment. Fig. 1 b) shows that the strongest received signal level results for a specific switched antenna element of the antenna array at one of the four sector antennas of the RSU. This antenna element can vary depending on the vehicle's driving trajectory or the radio environment. Fig. 1 c), on the other hand, shows that even for the reception of messages coming from the RSU, the strongest signal level can only be measured at a specific switched antenna element.

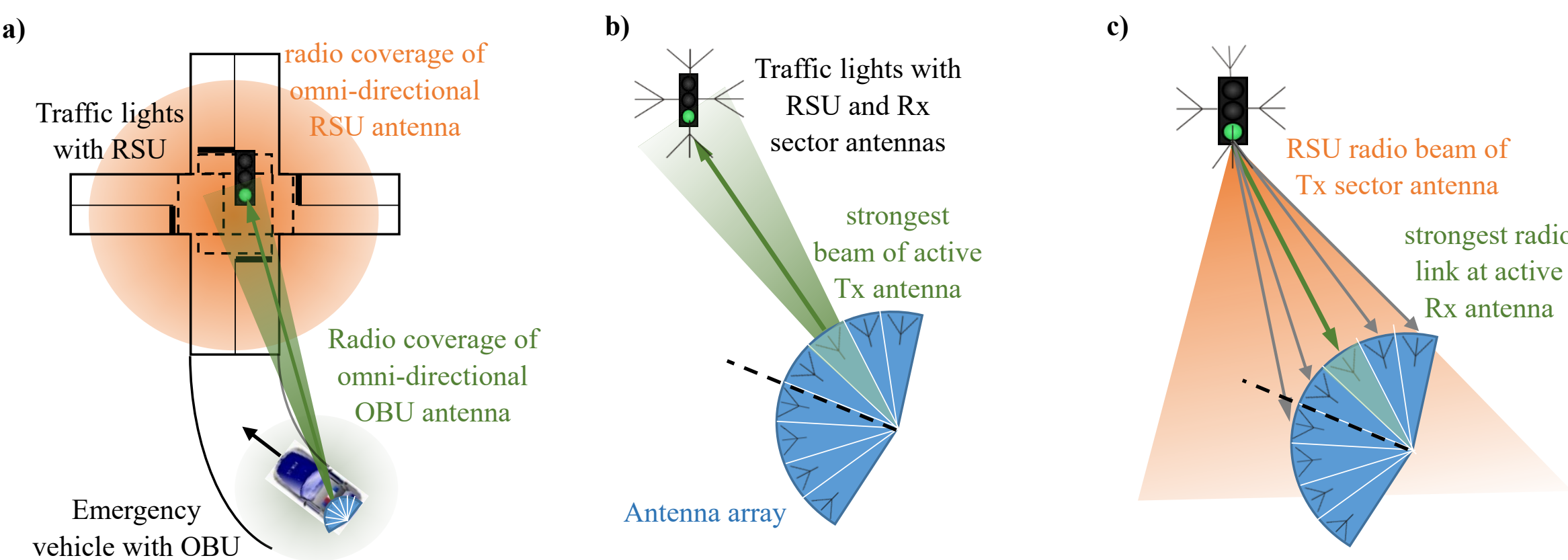

**Figure 1 - Basic concept of V2X communication scenario with OBU and RSU and switchable antenna array**

In the previous scientific works, beam-switching antennas and antenna arrays are discussed. In [14] especially the design of horn antenna arrays for different frequency ranges is considered. In [15], compiling diverse literature sources, several 360° beam-steering antennas are classified in terms of achieved antenna gain, antenna type, beam angle or the control method used. The work in [16] focuses specifically on beam design for beam switching at 60 GHz millimeter wave carrier frequency for V2X applications.
To implement the antenna concept, relevant radio parameters of the antenna apparatus to be developed were first defined. In addition to optimization in the frequency range of $f$ = 5.9 GHz for V2X applications according to ETSI ITS-G5, an antenna gain of $G \geq 10$ dBi was targeted. The single horn antenna was extended into an antenna array consisting of eight elements. In azimuth plane, an angular range of $\theta = \pm 60°$ is covered, with each antenna element covering an aperture angle of 15°. The respective antenna elements are controlled by an integrated solid-state radio frequency (RF) switch [17]. This is characterized by a high isolation of 30 dB and a comparatively low insertion loss of 2.5 dB at 6 GHz. The switching time is specified as 50 to 150 ns. The current consumption is between 5 and 9 mA at Vdd = 5 V. Control is via a 9-pin D-sub connector. In addition, the overall design has an omnidirectional sleeve antenna for continuous transmit and receive operation in the frequency range of $f$ = 5.9 GHz with a separate SMA connector and an antenna gain of $G$ = 2 dBi.
The overall construction consisting of antenna array, switch and omnidirectional antenna is equipped with three shear threads so that it can be installed on a mobile test vehicle. The horn antenna elements are covered as a radome. Fig. 2 shows the CAD model and the antenna construction manufactured by the company IRK – Dresden, Germany [18], on which the individual horn antenna elements are marked and numbered.

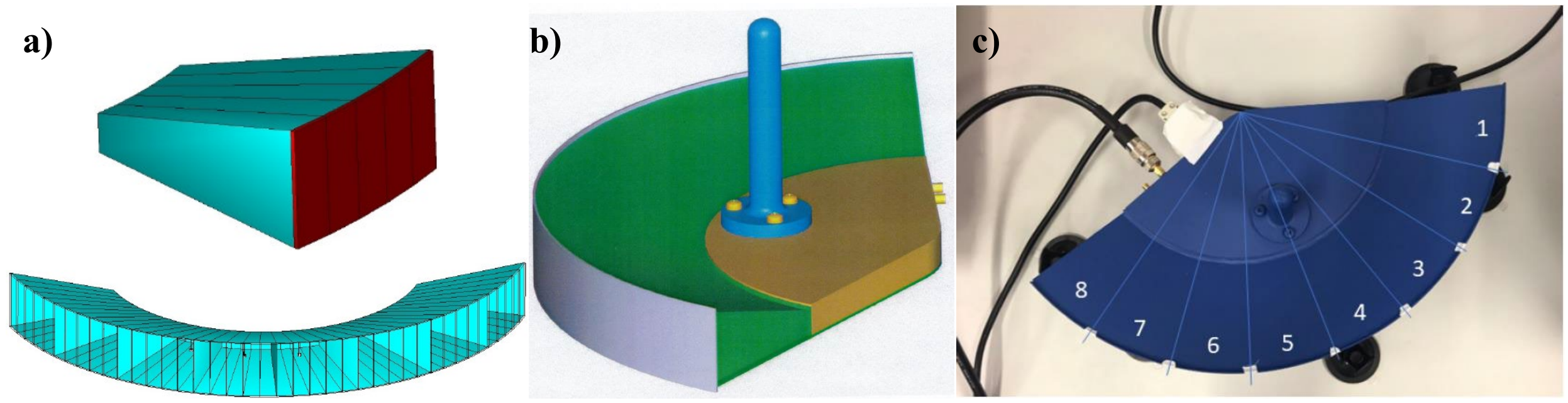

**Figure 2 –a) Simulation layout CAD horn ant., b) CAD ant. Apparatus, c) manufactured ant. apparatus**

The simulated antenna pattern of the 8 horn antenna elements with simulated radiation at $f$ = 5.9 GHz is shown in Fig. 3 a). This shows that the gain of the individual array antennas is $G$ > 11 dBi. The half-power width is given as ± 20° ... 25° in each case. This results in a high directional resolution of the antenna array. According to the data sheet, the simulated isolation between adjacent antenna elements ($S_{32}$ and $S_{34}$) is less than 50 dB [18]. The measured return loss of the individual antennas is shown in Fig. 3 b).

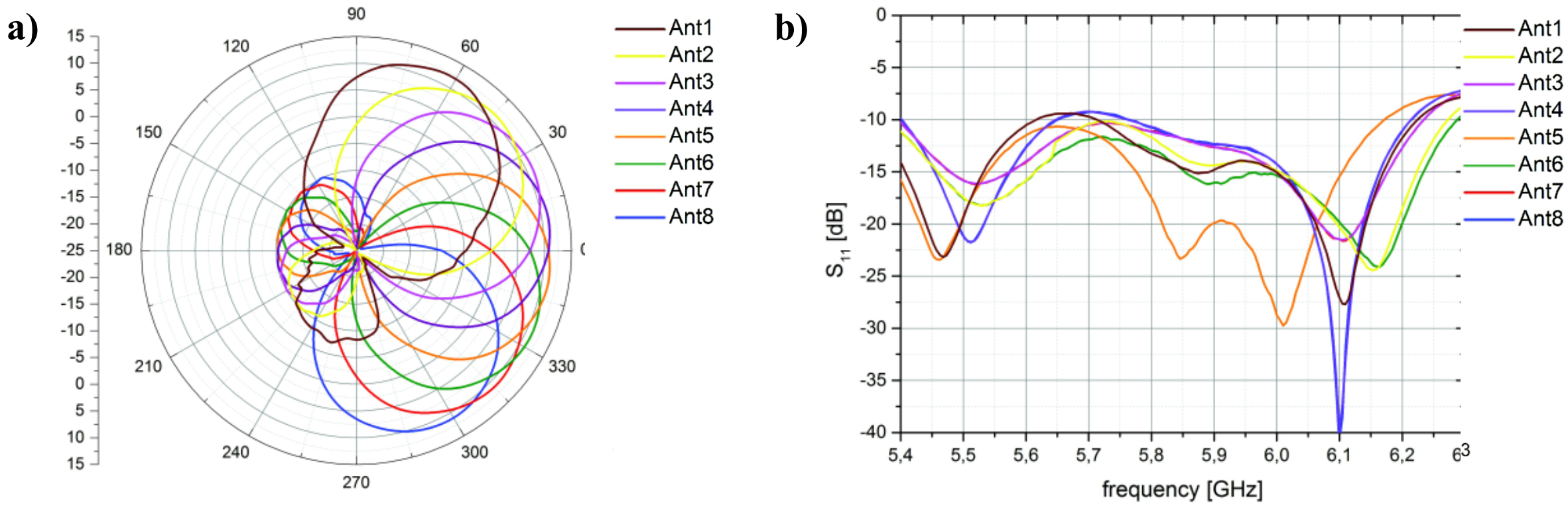

**Figure 3 – a) Simulated antenna beam pattern, $\theta$ = 0°, b) measured return loss**

## 3. Simulative and experimental results

### *3.1. Anechoic chamber measurements*

The evaluation of the manufactured antenna apparatus was initially carried out in a laboratory environment, an anechoic chamber. This allows electromagnetic influences on the measurement, e.g. due to multipath effects or signal interference, to be largely avoided. The measurements performed were intended to determine both directivity and antenna gain in the total radiation range of 120°. In the measurement setup, the antenna array as a transmitting (Tx) antenna and a receiving (Rx) antenna were placed at a distance $d$ = 4.70 m within the anechoic chamber. The gain $G_T$ of all antenna elements $n$ = {1, ..., 8} is determined in each case for a rotation angle $\theta$ around the antenna vertical axis by measuring $P_R$ and rearranging formula (1) with known other parameters. The antenna array was rotated in the range of $\theta$ = ± 60° in steps of 5° and all 8 antenna elements were switched sequentially. The Tx antenna array element was fed by a signal generator [19] with a CW signal of $f$ = 5.9 GHz. The RF antenna switch was controlled and powered by a Raspberry Pi Model B controller via GPIO. $P_R$ was measured using a spectrum analyzer [20]. Cable and link losses were previously determined by measurements and considered in the results. The frequency-specific antenna gain of the Rx antenna used was also considered in the calculation. The measurement setup in an anechoic chamber is shown in Fig. 4.

The result of the laboratory-based measurements in Fig. 5 shows that different antenna gains are obtained for all antennas of the array depending on the angle of rotation. The measured values range from $G_T$ = +11 dBi (Antenna 4) when aligned directly with the Rx antenna to $G_T$ = -21 dBi (Antenna 1) at an angle of 90° with respect to the antenna main lobe direction. Furthermore, it can be seen that for each of the antennas, the comparatively highest $G_T$ can be calculated in a particular angular segment. These angular segments are approximately in the range between 15 to 20°. A more precise quantification was not possible in the measurement setup due to the systemic limited rotation angle discretization interval. The horn antenna array thus achieves $G_T$ > 10 dBi over the entire horizontal coverage range of 120° considered.

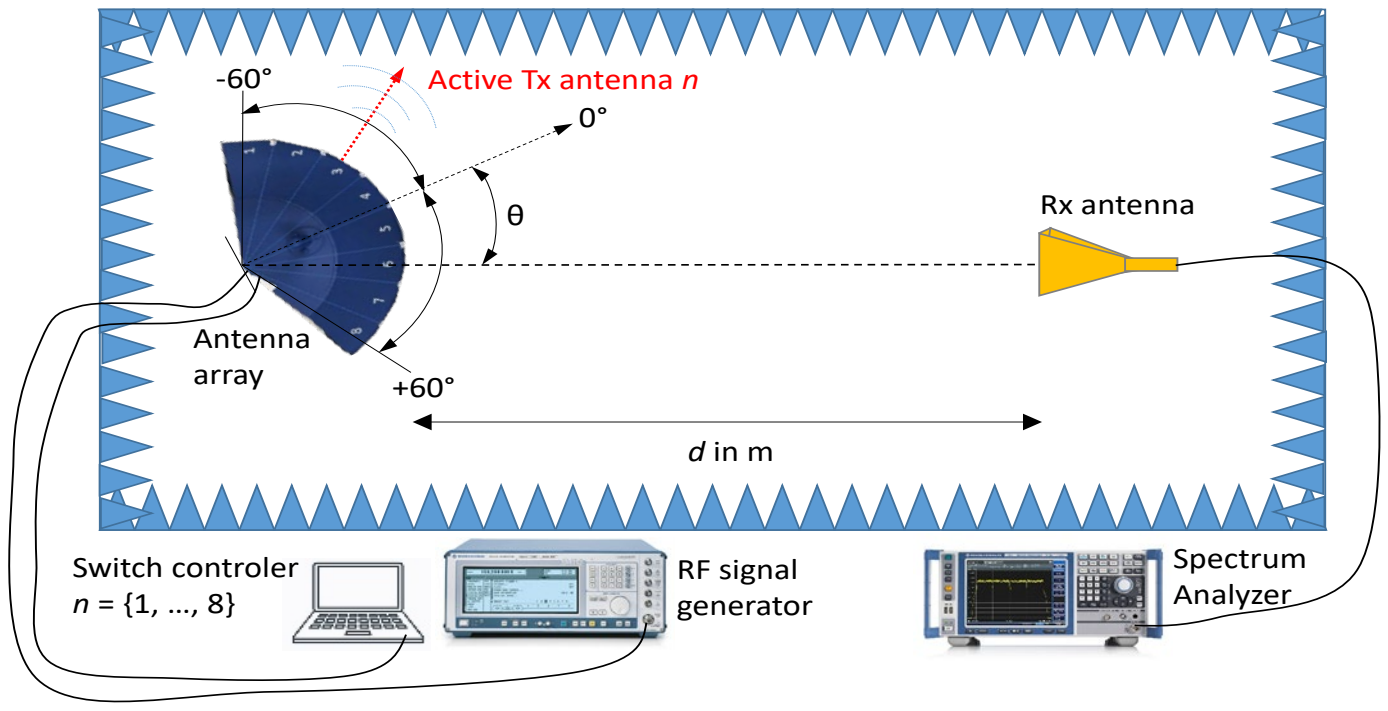

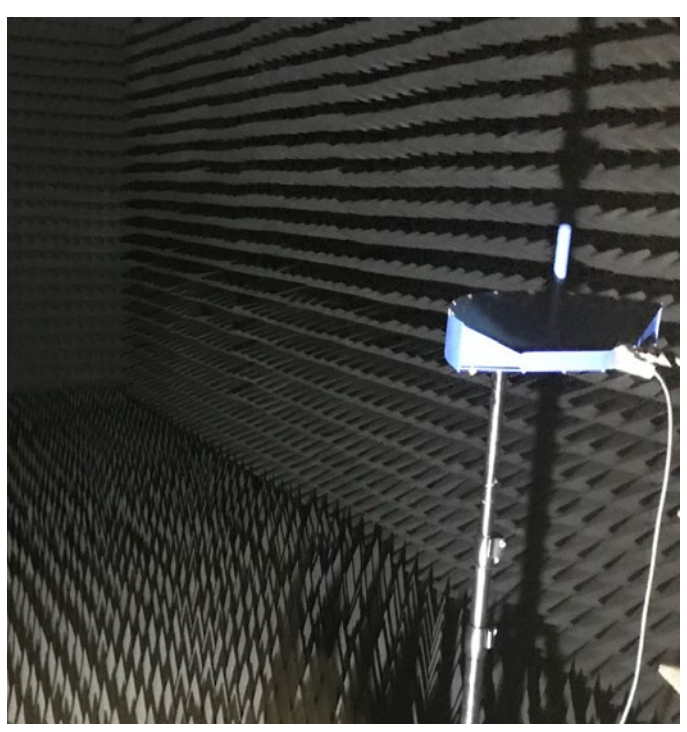

**Figure 4 – Laboratory measurement setup in the anechoic chamber**

This is due to the sequential switching of the antenna elements during the antenna array rotation around its vertical axis. The measurements thus confirm the values calculated in the simulation of the antenna pattern. Furthermore, the sidelobes of the antenna elements shown in Fig. 3 a) could also be observed in the measurements. E.g., considering Ant. 1, this is evident in an increasing signal level from $\theta > 30°$, although its main lobe moves away from the alignment axis of the Rx antenna as the angle of rotation increases.

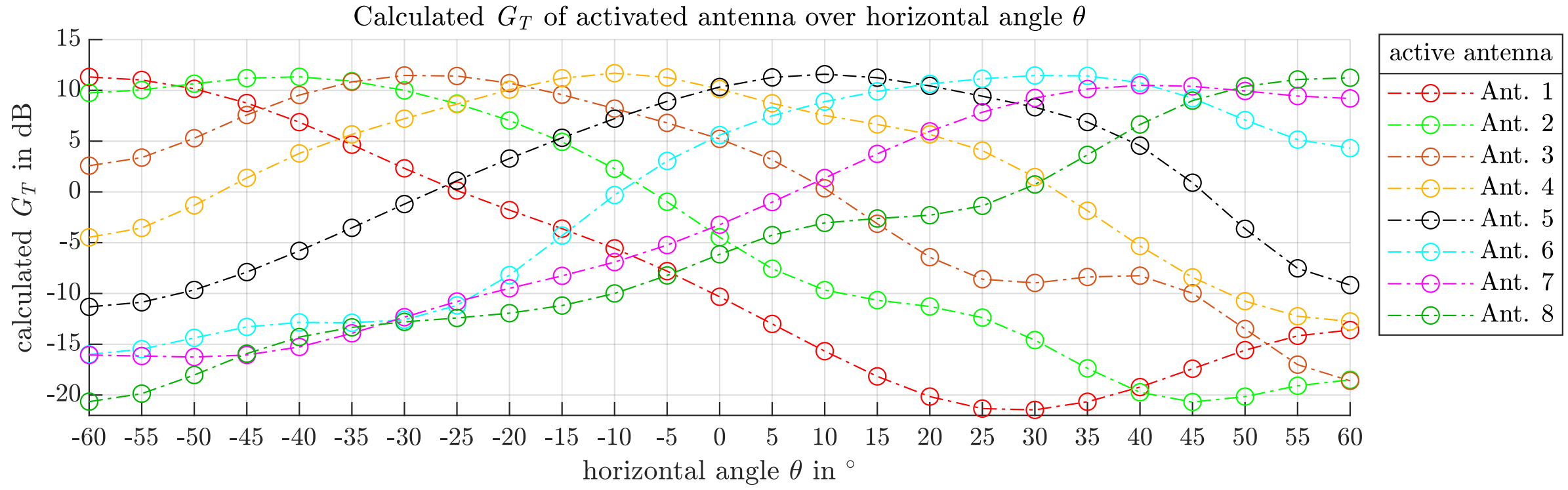

**Figure 5 – Calculated antenna gain $G_T$ of activated array antennas over horizontal rotation angle $\theta$**

*3.2 Field test simulation and measurements*

The field test investigations were carried out on the test field for automated driving at the University of Applied Sciences (HTW) in Dresden, Germany, which is characterized by ideal space and measurement conditions in an urban environment both for field test measurements and as a simulation environment. To evaluate the antenna apparatus in the field test, CW measurements were first carried out. The measurement setup, shown in Fig. 6, is similar to the laboratory-based investigations. For this setup, the antenna array was mounted on the roof of a research vehicle with $\theta = 0°$ aligned to the vehicle longitudinal axis. Onboard the vehicle the switch controller for switching the antenna elements and the signal generator were placed on the Tx side. The signal generator transmits CW signals with a transmission power of $P_T = 0$ dBm and $f = 5.9$ GHz. The signals received at the Rx antenna with $G_R = 16$ dBi were analyzed using a spectrum analyzer. The urban test environment offers the opportunity to perform measurement runs in the direction of the Rx antenna over a maximum distance $d = 127$ m under LOS conditions. $P_R$ measurements were performed at defined distances for all eight array antennas and the omnidirectional antenna.

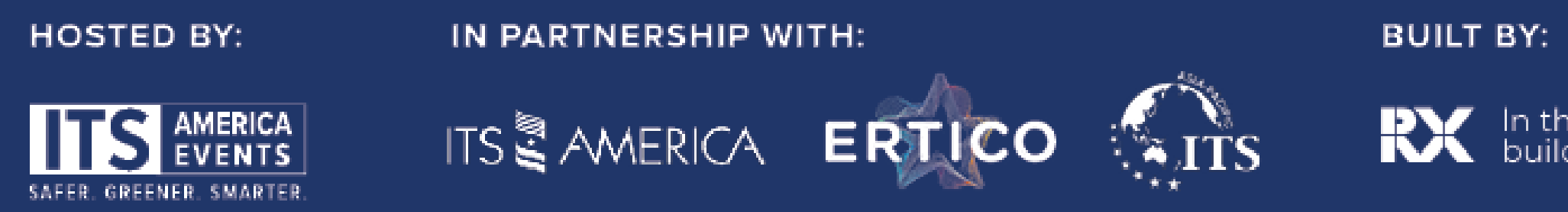

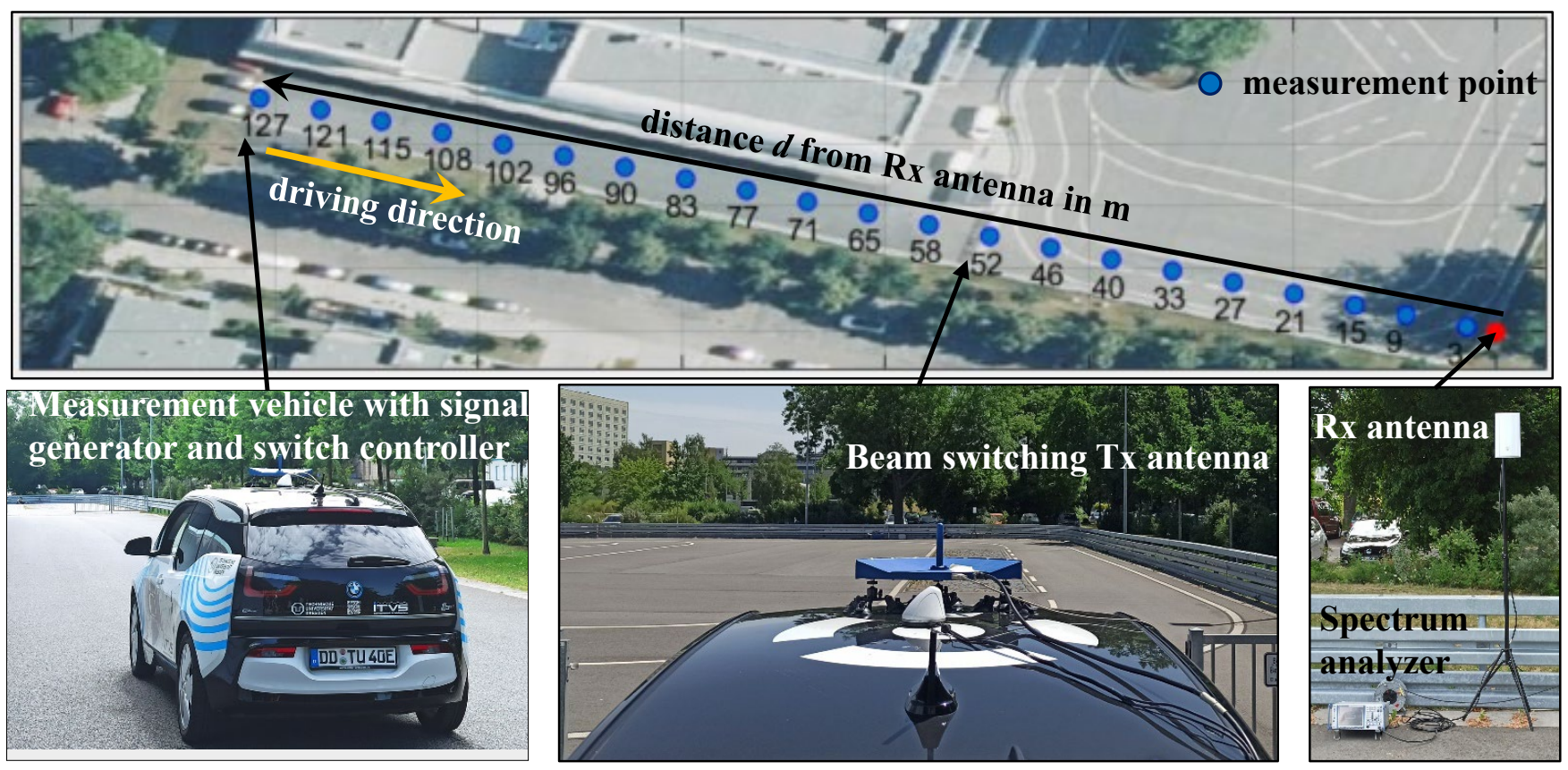

**Figure 6 – CW measurement setup in the test field with Tx antenna apparatus mounted on the vehicle roof**

The radio planning simulations were performed using the software Altair WinProp [21]. First, the test field environment is modeled as a CAD model. The antenna pattern used in the radio simulation was modeled as a single sector antenna with appropriate characteristics of maximum antenna gain, beam width and frequency range. With eight of these sector antennas in appropriate angular arrangement as shown in Fig. 3 a), the antenna array was created as a Tx antenna so that each of these antennas can be considered separately. In addition, the sleeve antenna on top of the antenna construction is modeled. The simulation model used to calculate the radio propagation is the Dominant Path Model (DPM) according to [22]. Between the transmitting antenna and the receiving antenna, only the most relevant signal path is considered with respect to the signal energy. The receiver level is located at a height of 1.80 m, just like the transmitting antenna. Receive level values are calculated in a grid of 1x1 m.
In addition to the CW measurements, a radio propagation simulation with DPM was performed for each of the measurement points and each antenna, and the $P_R$ levels at the location of the Rx antenna were determined. As an example for antennas 4 and 8, the simulation results in the CAD modeled test field for $d$ = 121 m are shown in Fig 7. Significantly higher $P_R$ can be observed for the Tx antenna elements aligned with the Rx antenna position (antenna 4). However, Tx antenna elements with the main lobe not pointing in the direction of vehicle longitudinal axis can provide stronger $P_R$ to the left or right of the vehicle (antenna 8 to the right). Depending on the location of the Rx antenna, for example a RSU site at a traffic light controlled intersection, optimum Tx antenna elements can thus be selected.

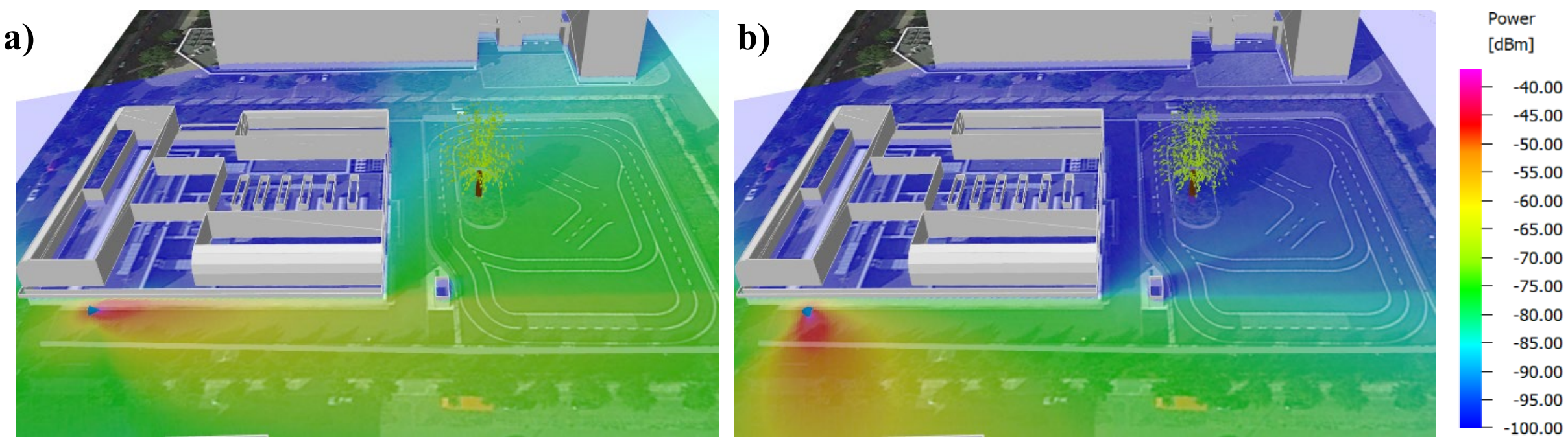

**Figure 7 – Radio propagation simulation in the CAD test field model for a) ant. 4 (straight), b) ant. 8 (right)**

Fig. 8 shows $P_R$ for the simulation and measurement over the distance to the Rx antenna and all antennas used. For the Tx antenna elements whose radio main lobes point in the direction of the Rx antenna, the highest $P_R$ can be observed over the entire distance (antennas 4 and 5). The greater $\theta$ of the antenna main lobes deviates from 0°, the lower $P_R$ at the same measurement point. Furthermore, the results show that $P_R$ can be significantly increased by an average of 8 dB with optimally aligned directional antennas compared to the omnidirectional antenna. Thus, according to chapter 2.1, an increase in the communication distance by a factor of approx. 2.5 can be achieved compared to the omnidirectional reference antenna. Less optimally aligned directional antennas, on the other hand, result in significantly lower $P_R$ compared with the omnidirectional antenna. This leads to the conclusion that, when using the smart antenna array for a V2X scenario, it is necessary to determine the best antenna element at each measurement point along a trajectory in order to keep $P_R$ as high as possible.

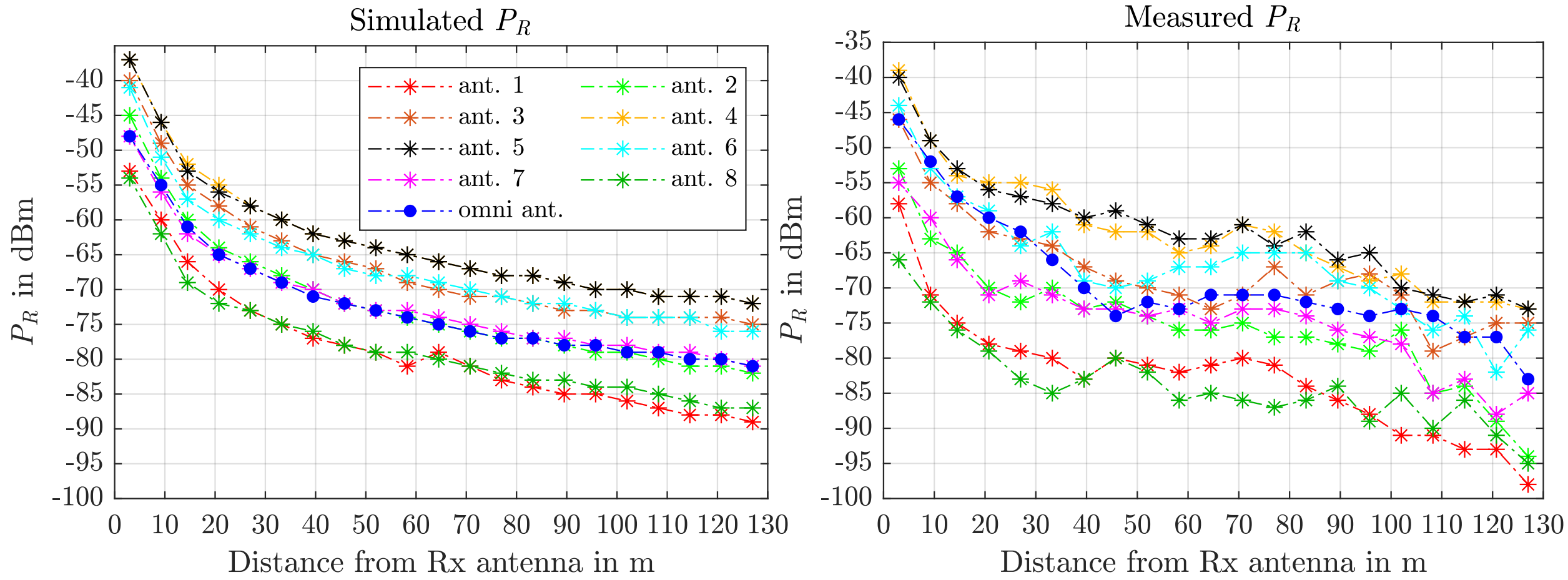

**Figure 8 – Simulated and CW measured $P_R$ over distance for all antenna array elements and omni antenna**

## 4. Applied Beam Switching setup for V2X services

The applicability of the considered Smart Antenna Array for the application scenario discussed in this paper, namely the prioritization of emergency vehicles at traffic signals in general and the increase of the communication distance in particular, was evaluated in a test setup on the test field. For this purpose, the beam-switching antenna mounted on the vehicle roof was connected to a OBU and an omnidirectional antenna to a RSU, both Cohda Wireless MK5 modules [23]. The bidirectional communication is based on the ETSI ITS-G5 standard and V2X messages were sent out - cooperative awareness messages (CAM) and DENM with the OBU, CAM, MAP and signal phase and timing messages (SPaT) with the RSU.
The GPS receiver in the OBU and RSU is used to determine not only the position but also the heading of the vehicle. This information is transmitted in the aforementioned message types via the communication channel and is thus available to the communication partner. Based on own position, heading, external position as well as the known antenna pattern of the antenna array, the calculation of an optimally aligned antenna element of the array is possible via geometric angle relations. An algorithm for automatic switching of the antenna elements has been implemented. Antenna switching is activated as soon as the geometric relationship of the line between Tx and Rx position with respect to the vehicle longitudinal axis is within the range $\theta = \{-100°, ..., 100°\}$. Otherwise, the integrated omnidirectional antenna is activated. On the one hand this ensures that switching is only activated when the vehicle is moving in the direction of the RSU, and on the other hand that the antenna pattern in this angular range has a higher antenna gain compared to the omnidirectional antenna. The trajectory of the measurement vehicle with respect to the RSU position and the measurement setup are shown in Fig. 9 a). Each measurement point is numbered from 1 to 195 and the active antenna is marked in color. The signal reception level of the OBU is reported as a hardware-specific Received Signal Strength Indicator (RSSI), which deviates by about 8 dB from the equivalent CW $P_R$ at the measurement point according to investigations in [3]. A radio propagation simulation carried out for this scenario, measurement points and antennas resulted in an offset of approximately 9 dB. The level measurement values of the RSSI are shown in Fig. 9 b), with the switched antenna assigned in color to each measurement value. In addition, a second measurement run was performed on the same trajectory, during which only the omnidirectional antenna was activated. In Fig. 9 c) the signal level difference of both measurement runs is given. According to this, a continuously higher received signal strength (RSS) of 6 dB on average can be observed due to the automatic switching of the antenna elements during the measurement run compared to a measurement run with omni-antenna. It is higher in areas where the vehicle longitudinal axis points in the direction of the RSU than in areas where the vehicle moves orthogonally to the RSU. This is because the RSU is outside the optimum effective coverage area of the antenna with the highest antenna gain. In summary, the scenario with V2X radio modules and automated antenna beam switching has demonstrated that the antenna apparatus enables higher RSSI within an angle $\theta = \pm 110°$ in the vehicle longitudinal axis compared to an omnidirectional antenna.

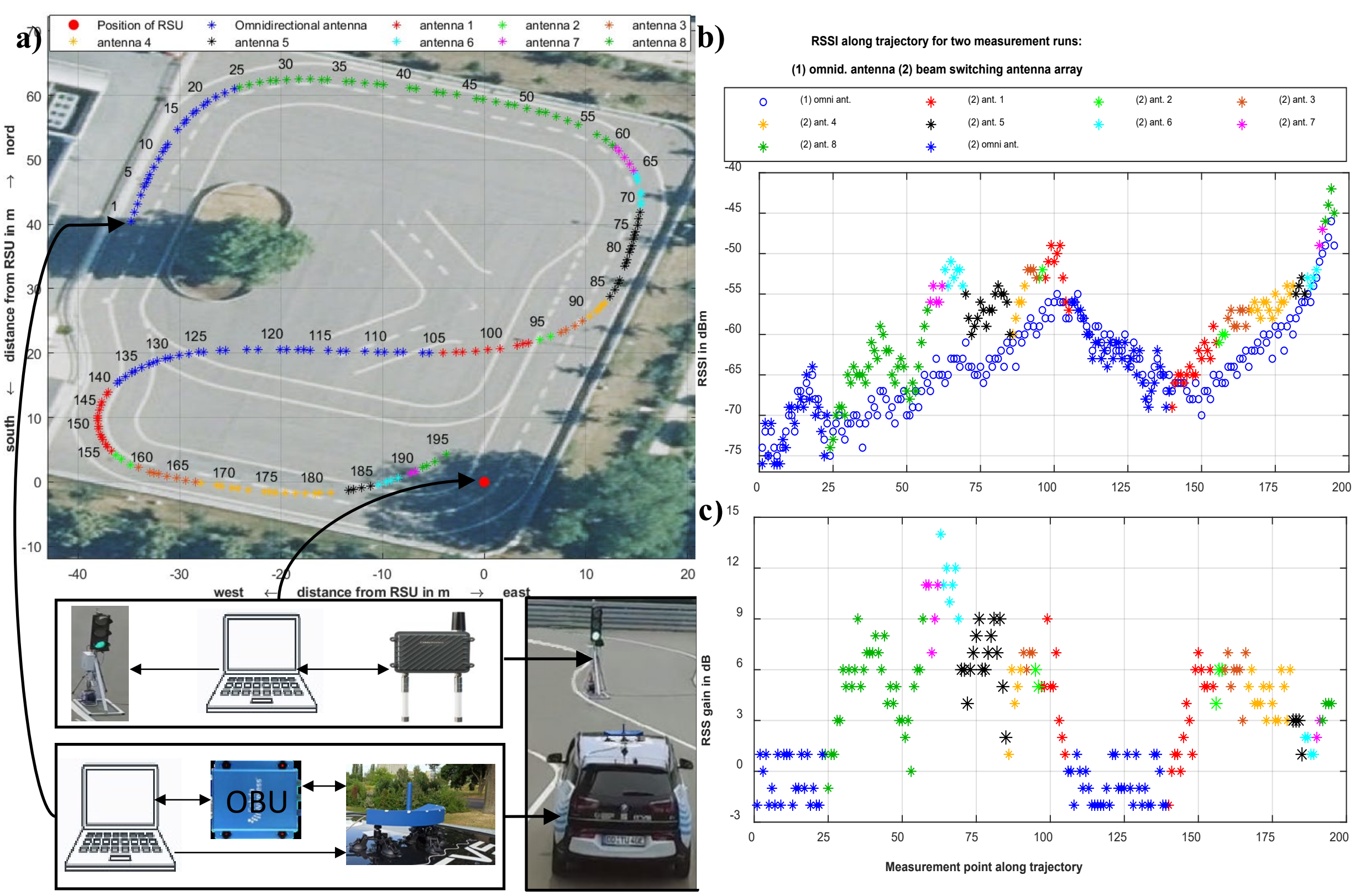

**Figure 9 – a) V2X measurement setup and trajectory in the test field, b) RSSI measurements for two measurement runs alsong the trajectory with (1) omni antenna, (2) beam switching antenna and algorithm, c) RSS gain of antenna array comparing the two measurement runs**

**5. Conclusion and Outlook**

As a result of this research, a switchable antenna array was evaluated for optimized radio coverage in a traffic telematics environment. In this context, an antenna construction was designed and implemented specifically for the requirements of a V2X scenario. Subsequent investigations in a laboratory environment, in a radio planning simulation and in a traffic telematics test field environment showed that the directional antennas arranged in an array and individually switchable have a high antenna gain over a horizontal angle of at least 120°. As a result, an increase in communication robustness can be demonstrated by increasing the signal reception level compared to conventional omnidirectional antennas. The resulting increased communication distance by a factor of approx. 2.5 is of particular relevance for the use case considered in the traffic sector of giving priority to emergency vehicles at traffic lights. An equivalent scenario with OBU and RSU for service measurements in the V2X communication standard ETSI ITS-G5 with $f$ = 5.9 GHz could be carried out in a test field for automated driving. With regard to the resulting signal reception levels, this again demonstrated the beneficial use of the automated beam switching antenna array based on georeferenced control. Furthermore, by using solid-state switches to switch the antenna elements, a low switching latency required for safety-critical applications could potentially be achieved. Investigations to prove and quantitatively estimate the low switching latency are to be carried out in further work. Furthermore, the use of such antenna systems for other areas of application and means of transport is conceivable and should be investigated. These include, for example, public transport systems such as buses, tramways or trains. It may also be worth investigating the use of frequency-matched switchable antenna arrays for transmission technologies currently in widespread use or in the process of standardization, e.g., in the field of 5G/6G for future communications applications.

**Acknowledgement**

The research was part of the 'fast sign' project (FKZ: 03ZZ0527C) and funded by the German Federal Ministry of Education and Research within the research program 'Zwanzig20 – Partnerschaft für Innovation'.

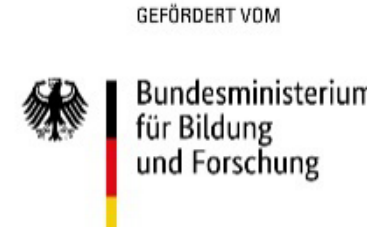

**References**

1. Bey, T., Tewolde, G. (2019). Evaluation of DSRC and LTE for V2X. *2019 IEEE 9th Annual Computing and Communication Workshop and Conference (CCWC)*, 2019, pp. 1032-1035.

2. Younes, M.B., Boukerche, A. (2018). An efficient dynamic traffic light scheduling algorithm considering emergency vehicles for intelligent transportation systems. *Wireless Netw* 24**,** 2451–2463 (2018).

3. Ußler, H., Setzefand, C., Kanis, D., Schwarzbach, P., Michler, O*.:* Efficient coverage planning for full-area C-ITS communications based on radio propagation simulation and measurement tools. In Proc. *27th ITS World Congress*, Hamburg. ERTICO (ITS Europe)

4. ETSI EN 302 663 (2020). Intelligent Transport Systems (ITS); ITS-G5 Access layer specification for Intelligent Transport Systems operating in the 5 GHz frequency band, V1.3.1, European Standard

5. Saunders, S.R., Aragón-Zavala, A. (2007). Antennas and Propagation for Wireless Communication Systems; Hoboken, NJ, USA: Wiley.

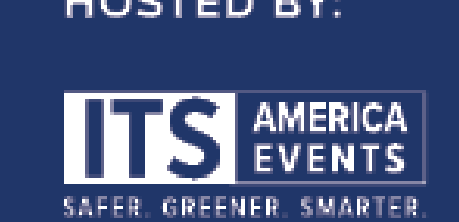

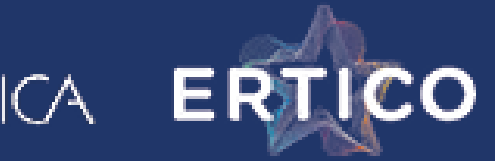

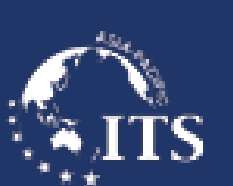

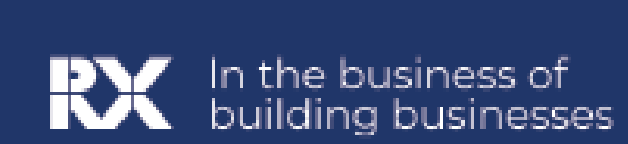

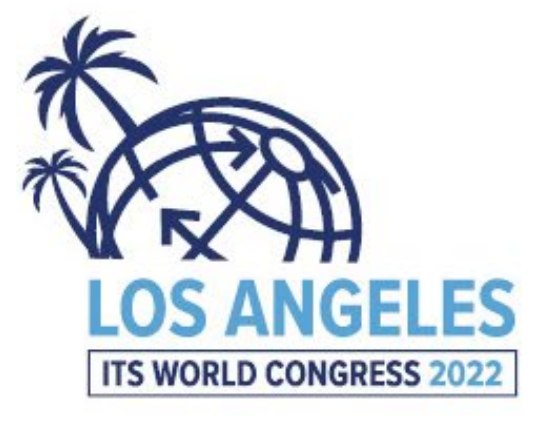

6. Salous, S. (2013). Radio Propagation Measurement and Channel Modelling; John Wiley & Sons Ltd, 2013

7. Balanis, C. A. (2015). *Antenna theory: analysis and design*. John Wiley & Sons Ltd, 2015.

8. Schwarzbach, P., Engelbrecht, J., Michler, A., Schultz, M., Michler, O. (2020). Evaluation of Technology-Supported Distance Measuring to Ensure Safe Aircraft Boarding during COVID-19 Pandemic. *Sustainability* 2020, 12, 8724.

9. Ußler, H., Setzefand, S., Michler, O. (2022). An empirical study on V2X radio coverage using leaky coaxial cables in road crash barriers. Proceedings of the *24st EURO Working Group on Transportation Meeting*, EWGT 2021. Aveiro, Portugal.

10. Jung, A., Schwarzbach, P., Michler, O. (2020). Future Parking Applications: Wireless Sensor Network Positioning for Highly Automated In-House Parking, Proceedings of the *17th Intern. Conf. ICINCO*, Paris, 710-717.

11. Shivapanchakshari, T.G., Aravinda, H.S. (2017). Review of Research Techniques to Improve System Performance of Smart Antenna. *Open Journal of Antennas and Propagation*, 5, 83-98. 2017.

12. Jain, M., Agarwal, R.P. (2016). Capacity & Coverage Enhancement of Wireless Communication Using Smart Antenna System. *2nd International Conference on Advances in Electrical, Electronics, Information, Communication and Bio-Informatics (AEEICB)*. Chennai. 310-313.

13. Arsad, A.Z., Sebastian, G., Hannan, M.A., Ker, P.J., Rahman, M.S.A., Mansor, M., Lipu, M.S.H. 2021. Solid State Switching Control Methods: A Bibliometric Analysis for Future Directions. *Electronics* 2021, *10*, 1944.

14. Pranonsatit S., Holmes, A. S., Lucyszyn, S. 2010. Sectorised horn antenna array using an RF MEMS rotary switch. *2010 Asia-Pacific Microwave Conference*, 2010, pp. 1909-1913.

15. Yang, Y., Zhu, X. (2018). A Wideband Reconfigurable Antenna With 360° Beam Steering for 802.11ac WLAN Applications. In *IEEE Transactions on Antennas and Propagation*, vol. 66, no. 2, pp. 600-608.

16. Va, V., Shimizu, T., Bansal, G., Heath, R. W. (2016). Beam design for beam switching based millimeter wave vehicle-to-infrastructure communications. In *2016 IEEE International Conference on Communications (ICC)*, 2016, pp. 1-6.

17. Analog Devices, Inc. HMC321LP4 / 321LP4E. GaAs MMIC SP8T NON-REFLECTIVE POSITIVE CONTROL SWITCH, DC* - 8 GHz

18. IRK – Dresden, microwave engineering and antenna development. (2019). Simulations und Messbericht. Entwicklung und Aufbau eines 5,9 GHz Antennen-Array. 05.04.2019

19. Rohde & Schwarz GmbH & Co. KG, Vector Signal Generator, SMIQ06B, 1125.5555.06

20. Rohde & Schwarz GmbH & Co. KG (2015). FSVR Real-Time Spectrum Analyzer 7, Specifications, PD 5214.3381.22, Version 03.00

21. Altair Engineering, Inc.: Altair Feko™-Suite. Altair WinProp 2020.

22. Wahl, R., Wolfle, G. (2006). Combined urban and indoor network planning using the dominant path propagation model. *First European Conference on Antennas and Propagation*. Nice, 2006, pp. 1-6.

23. De Haaij, D., Strauss, U., Sloman, M. (2017). Cohda Wireless MK5 OBU Specification, Reference: CWD-P0052-OBU-SPEC-WW01-186, Version 1.6, 31th Oct 2017

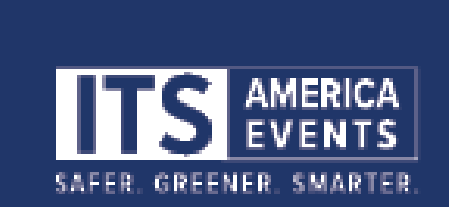